\documentclass[prl,aps,twocolumn,preprintnumbers,nofootinbib,notitlepage]{revtex4}
\usepackage{amssymb, amsthm}
\usepackage{amsmath}
\usepackage{graphicx}% Include figure files
\usepackage{hyperref}
\usepackage{dcolumn}% Align table columns on decimal point
\usepackage{bm}%
\usepackage{float} 
\usepackage{soul}
\usepackage{xcolor}
\begin{document}

\title{Axionless strong CP problem solution: the spontaneous CP  violation case}% 

\author{Rodolfo Ferro-Hernández$^1$, Stefano Morisi$^{2,3}$ and Eduardo Peinado$^4$$^{,5}$ }\affiliation{$^1${\rm PRISMA}$^+$ Cluster of Excellence and 
Institute for Nuclear Physics, Johannes Gutenberg-University, 55099 Mainz, Germany\\
$^2$INFN - Sezione di Napoli, Complesso Univ. Monte S. Angelo, I-80126 Napoli, Italy\\
$^3$Dipartimento di Fisica ``Ettore Pancini'', Università degli studi di Napoli ``Federico II'', Complesso Univ. Monte S. Angelo, I-80126 Napoli, Italy\\
$^4$Instituto de F\'{\i}sica, Universidad Nacional Aut\'onoma de M\'exico, A.P. 20-364, Ciudad de M\'exico 01000, M\'exico.\\
$^5$Departamento de F\'isica, Centro de Investigaci\'on y de Estudios Avanzados del Instituto Polit\'ecnico Nacional.}

%\date{\today}% It is always \today, today, but any date may be explicitly specified

%\keywords{Neutron EDM, strong CP problem}%Use showkeys class option if keyword display desired
%\preprint{}
  \begin{abstract}
 We propose an alternative to the axion mechanism for addressing the charge parity (CP) problem in quantum chromodynamics (QCD). Our approach involves imposing CP as an inherent symmetry of the Lagrangian, which is then spontaneously broken. To generate the correct texture for the Yukawa matrices, we introduce a discrete $\mathbb{Z}_2$
  symmetry that is softly broken by the scalar potential. By identifying a benchmark point for the Yukawa couplings that aligns with the measured quark masses, the CKM matrix, and low-energy flavor-changing constraints, our findings suggest that this model offers a viable solution to the CP problem.

  \end{abstract}
  \maketitle
 \section{Introduction}
The non-observation of the neutron’s dipole moment 
%$|d_n|< 3   10^{-26}e\,cm$ 
provides a strong limit on the CP-violating parameter 
\begin{equation}\label{theta}
\bar{\theta} = \theta + \text{Arg Det}(M_u\,M_d),
\end{equation}  
where \(\theta\) is the coefficient of the QCD topological term 
\begin{equation}
\mathcal{L}_{\rm QCD} \supset \frac{\theta}{32 \pi^2} G^{\mu\nu}\tilde{G}_{\mu\nu},
\end{equation}  
\(G\) is the gluon field strength, and \(M_{u,d}\) are the quark mass matrices. The strong CP problem consists of explaining the extreme smallness of the parameter \(\bar{\theta} \lesssim 10^{-10}\) (see \cite{DiLuzio:2020wdo} for a recent review). The strategy to solve this problem involves implementing symmetries (continuous or discrete) that constrain \(\bar{\theta}\).

Peccei and Quinn proposed a solution based on a continuous global symmetry \(U(1)_{PQ}\) \cite{Peccei:1977hh} implying the existence of a pseudo-goldstone boson (the axion) that is also a viable dark matter candidate \cite{Abbott:1982af,Preskill:1982cy,Dine:1982ah}. Even though the Peccei-Quinn mechanism is elegant and simple, it suffers from the so-called quality problem. Furthermore, Global symmetries are broken by gravity, meaning that higher-dimensional operators, suppressed by powers of the Planck scale \cite{Kamionkowski:1992mf,Holman:1992us,Barr:1992qq}, resulting in a minimum of the potential at \(\bar{\theta}\neq 0\), unless the coefficients of these operators are exponentially suppressed (\(\lesssim 10^{-55}\)).

%Therefore, the Peccei-Quinn mechanism introduced to solve a fine-tuning problem is subject to a similar situation, rendering this solution unnatural.

On the other hand, in the so-called {\it axionless}  models relying on discrete global symmetries (see Refs.~\cite{Mohapatra:1978fy,Beg:1978mt,Georgi:1978xz,Goffin:1979mt,Nelson:1983zb,Barr:1984qx,Babu:1989rb} for pioneer works) the quality problem is absent. 
These models rely on imposing CP or P besides extending either the gauge symmetry or the quark sector to make both terms in eq.\,(\ref{theta}) vanish. The topological term proportional to ${\theta}$ is absent due to CP invariance,  and  $\mbox{Det}(M_u\,M_d)$ is a real quantity (at tree-level) even if $M_{u,d}$ are complex, allowing to obtain the sizable phase of the Cabibbo-Kobayashi-Maskawa (CKM) matrix \cite{ParticleDataGroup:2022pth} (a recent advance in this direction is given in Ref.~\cite{Vecchi:2014hpa}). Nevertheless, once the symmetry is broken either spontaneously or softly, contributions to $\overline{\theta}$ will typically come from one-loop diagrams. 
Recently, axionless models has been revamped \cite{Diaz-Cruz:2016pmm,Liao:2017vaf,Senjanovic:2020int,Babu:2023srr,Babu:2023dzz,Li:2024sln,Bonnefoy:2023afx,Camara:2023hhn,Murai:2024alz}, also in connection with the flavor problem \cite{Antusch:2013rla,Fong:2013sba,Agrawal:2017evu,Hall:2024xbd} and in particular by means of the modular invariance
\cite{Feruglio:2023uof,Petcov:2023vws,Higaki:2024jdk,Petcov:2024vph, Penedo:2024gtb,Feruglio:2024ytl}.
Different axionless solutions have also been proposed based on supersymmetry \cite{Hiller:2001qg,Nakagawa:2024ddd},
vacuum accumulation mechanism \cite{Dvali:2005zk},
extra dimensions \cite{Bezrukov:2008da},
anomalous $U(1)$ \cite{Hook:2014cda}, long distance vacuum effects \cite{Nakamura:2021meh}.
%\section{The Model}

This work considers an axionless solution to the strong CP problem based on Georgi's mechanism~\cite{Georgi:1978xz}. Georgi's original work focused on two families of quarks. Here, we extend the analysis to the realistic case with three families. We compare our results with current experimental constraints, including quark masses and mixings, low-energy flavor-changing neutral currents, and bounds on $\overline{\theta}$. By assuming that the heaviest scalar has a mass of $\lesssim1$ TeV, the strong constraints on the scalar sector from perturbative unitarity, perturbativity, and boundedness from below are included (see Ref.~\cite{Nebot:2019qvr}). We conclude that this model remains a viable solution to the CP problem.

 \section{The model}

The model contains two Higgs fields, $\phi_1$ and $\phi_2$, that are respectively even and odd under a $\mathbb{Z}_2$ symmetry. The third left-handed quark family is odd under the discrete $\mathbb{Z}_2$ symmetry and couples only with the second Higgs, $\phi_2$. The rest of the Standard Model (SM) fermions are even under $\mathbb{Z}_2$ and therefore couple only with the first Higgs, $\phi_1$.

If the scalar potential is completely invariant under $CP \times \mathbb{Z}_2$, the two Higgs fields take real vacuum expectation values (vevs). However, if the scalar potential contains a $\mathbb{Z}_2$ soft-breaking term, $\phi_1$ and $\phi_2$ develop complex vevs after CP is spontaneously broken, as shown in Ref.~\cite{Georgi:1978xz}.

The minimization of the two Higgs potential with a softly-broken $\mathbb{Z}_2$ symmetry has been recently studied, for instance, in Ref.~\cite{Nebot:2018nqn} (more references can be found in the review given in Ref.~\cite{Branco:2011iw}). Complex vevs are required to fit the CP-violating phase of the CKM matrix.

The key idea is that since every row in the Yukawa coupling matrix for the up-quarks and down-quarks come with $\phi_i$ and $\tilde{\phi}_{i} = -i \sigma_2 \phi_i^*$ respectively, the phases entering each mass matrix are opposite, and therefore the argument of $\text{Det}(M_u)$ exactly cancels the argument of $\text{Det}(M_d)$. Moreover, if $\phi_2$ couples only to the third left-handed quark family, the Flavor Changing Neutral Currents (FCNCs) are under control~\cite{Botella:2015hoa}.

As mentioned before, in the three-family realization of Georgi's mechanism considered here, all the right-handed quark fields, $U_{iR}$ and $D_{iR}$, as well as the left-handed quark doublets
\begin{equation}\label{matter}
Q_{1L} =
\left( 
\begin{array}{c}
U_{1L} \\
D_{1L}
\end{array}
\right)\,,\qquad
Q_{2L} =
\left( 
\begin{array}{c}
U_{2L} \\
D_{2L}
\end{array}
\right)\,,
\end{equation}
are even under $\mathbb{Z}_2$, while the left-handed quarks $Q_{3L}$ are odd under $\mathbb{Z}_2$. Therefore, the Yukawa couplings become
\begin{align}\label{lag}
    \mathcal{L}_Y &= \sum_{i=1}^2 \sum_{k=1}^3 Y^u_{ik} \overline{Q_{iL}} \, U_{kR} \, \phi_1 
    + \sum_{i=1}^3 Y^u_{3i} \overline{Q_{3L}} \, U_{iR} \, \phi_2 \nonumber \\
    &+ \sum_{i=1}^2 \sum_{k=1}^3 Y^d_{ik} \overline{Q_{iL}} \, D_{kR} \, \tilde{\phi}_1 
    + \sum_{i=1}^3 Y^d_{3i} \overline{Q_{3L}} \, D_{iR} \, \tilde{\phi}_2 + \text{h.c.}
\end{align}
The two Higgs fields take vevs $\langle \phi_i \rangle = (0,v_i e^{i \alpha_i})^T$, where $v_i$ are real and $\alpha_i$ are phases. Because the model imposes CP invariance, the 18 Yukawa couplings $Y^u_{ij}$ and $Y^d_{ij}$ are real; however, spontaneous CP breaking implies that $\alpha_i \ne 0$. It is possible to show that one of the phases $\alpha_i$ (we choose $\alpha_1$) can be reabsorbed, leaving only one phase $\alpha_2 \equiv \alpha$.

The two Higgs model gives (from the diagonalization of the most generic potential involving $\phi_1$ and $\phi_2$) two scalar (CP even) neutral states $h$ and $H$, one pseudoscalar (CP odd) state $A$ and two charged ones $H^{\pm}$.

 \section{Quark mass textures and radiative corrections}

\begin{figure}
    \centering
    \includegraphics[scale=0.25, angle=-90]{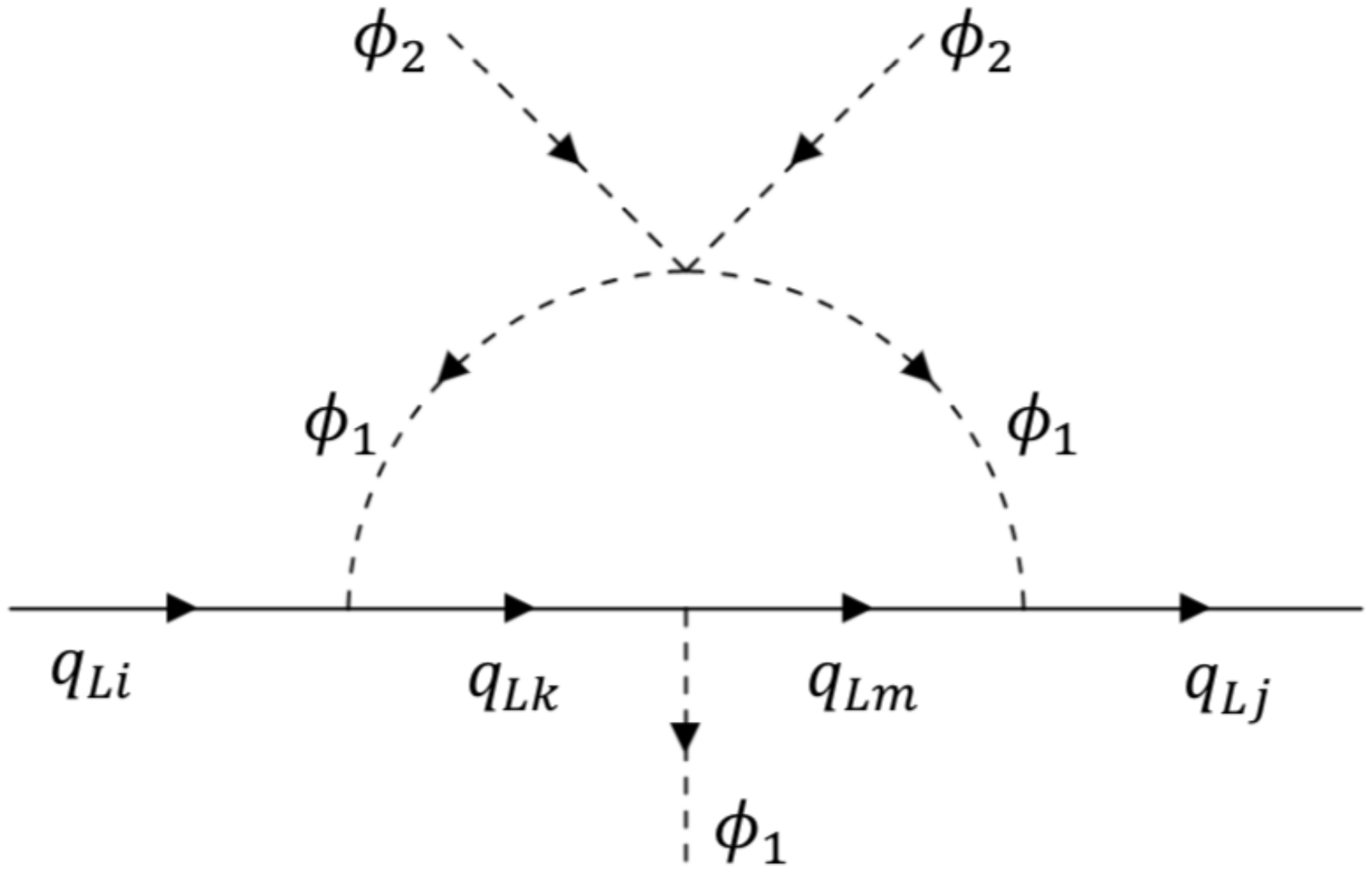}
        \includegraphics[scale=0.25, angle=-90]{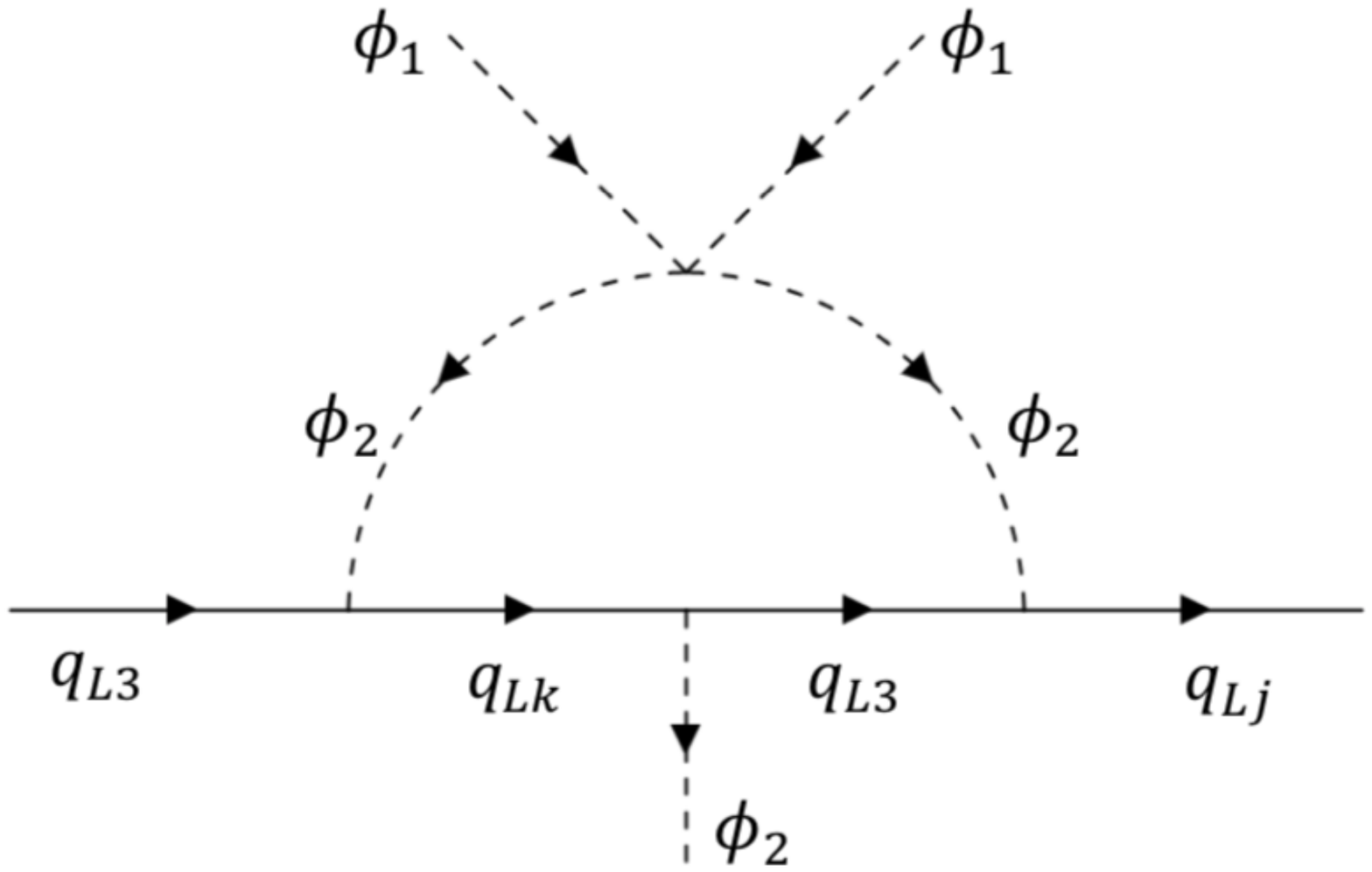}
    \caption{One-loop diagrams correcting $\bar{\theta}$.}
    \label{fig:diagram}
\end{figure}

After electroweak symmetry breaking, from (\ref{lag}) we obtain
\begin{eqnarray}
M_u^0&=&\frac{v_1}{\sqrt{2}}
\left( 
\begin{array}{ccc}
Y^u_{11} & Y^u_{12}& Y^u_{13}\\
Y^u_{21} & Y^u_{22}& Y^u_{23}\\
Y^u_{31} t_\beta e^{i\alpha} & Y^u_{32}  t_\beta e^{i\alpha} & Y^u_{33} t_\beta  e^{i\alpha} 
\end{array}
\right)\,,\nonumber\\
&& \label{Mud}\\
M_d^0&=&\frac{v_1}{\sqrt{2}}
\left( 
\begin{array}{ccc}
Y^d_{11} & Y^d_{12}& Y^d_{13}\\
Y^d_{21} & Y^d_{22}& Y^d_{23}\\
Y^d_{31} t_\beta e^{-i\alpha} & Y^d_{32}  t_\beta e^{-i\alpha} & Y^d_{33} t_\beta  e^{-i\alpha} 
\end{array}
\right)\nonumber
\end{eqnarray}
where $t_\beta\equiv \tan\beta=v_2/v_1$ and the 0-apex refers to tree-level mass matrices. Loop contributions are expected to modify the $M_{u,d}^0$ pattern. The quark mass matrices, 
including one-loop corrections (see the diagram of figure\,(\ref{fig:diagram})), take the form
$$
M_{u,d}=M_{u,d}^0+M_{u,d}^{(1)},
$$
where $(1)$ apex indicates the one-loop contribution that is
\begin{eqnarray}
M_{u}^{(1)}&\approx&
\left( 
\begin{array}{ccc}
\delta^{u}_{11}\, e^{+2i\alpha}  & \delta^{u}_{12}\, e^{+2i\alpha} & \delta^{u}_{13}\,e^{+2i\alpha} \\
\delta^{u}_{21}\, e^{+2i\alpha} & \delta^{u}_{22}\,e^{+2i\alpha} & \delta^{u}_{23}\,e^{+2i\alpha} \\
\delta^{u}_{31}\,t_\beta^{-1}\, e^{-i\alpha}& \delta^{u}_{32}\,t_\beta^{-1} \,e^{-i\alpha}&  \delta^{u}_{33} \,t_\beta^{-1}\, e^{-i\alpha}
\end{array}
\right)\,, \nonumber\\
&& \\
M_{d}^{(1)}&\approx&
\left( 
\begin{array}{ccc}
\delta^{d}_{11}\, e^{-2i\alpha}  & \delta^{d}_{12}\, e^{-2i\alpha} & \delta^{d}_{13}\, e^{-2i\alpha} \\
\delta^{d}_{21}\, e^{-2i\alpha} & \delta^{d}_{22}\,e^{-2i\alpha} & \delta^{d}_{23}\,e^{-2i\alpha} \\
\delta^{d}_{31}\,t_\beta^{-1}\, e^{+i\alpha}& \delta^{d}_{32}\,t_\beta^{-1} \,e^{+i\alpha}&  \delta^{d}_{33} \,t_\beta^{-1} \,e^{+i\alpha}
\end{array}
\right)\,,\nonumber
\end{eqnarray}
where
(for $i=1,2$ and $j=1,2,3$)
\begin{equation}
\delta^{u,d}_{ij} =\frac{ \lambda_5v_1^3 t_\beta^2}{16 \pi^2 M^2}  \sum_{k=1}^3 \sum_{m=1}^2 Y^{u,d}_{ik}\, {Y^{u,d}}^*_{mk} \,Y^{u,d}_{mj}\,, 
\end{equation}
and for $i=3$,
\begin{equation}
\delta^{u,d}_{3j}=\frac{ \lambda_5v_1^3 t_\beta^2}{16 \pi^2 M^2} \sum_{k=1}^3 {Y^{u,d}}_{3k} \,{Y^{u,d}}^*_{3k} \,Y^{u,d}_{3j}\,,
\end{equation}
where $M\sim 1$\,TeV is the scale of heaviest Higgs in the loop, $\lambda_5$ is the coupling of  $ ((\phi_1^\dagger \, \phi_2)^2 + (\phi_2^\dagger \, \phi_1)^2)$ term in the potential (see \cite{Nebot:2018nqn}).

%In the definition of $\delta$'s, we note that the conjugation is put for completeness even if not strictly necessary since the Yukawa couplings are real in this model.

Including one-loop contribution, a simple estimation gives 
\begin{equation}\label{loop}
    \mbox{Arg\,Det}(M_u\,M_d) \sim \sin \alpha \, \lambda_5 \, f_Y \,\frac{v_1^2 \tan\beta}{16 \pi^2 M^2}\sim \lambda_5 \, f_Y\, 10^{-5}\,,
\end{equation}
where $f_Y$ is a function of the $\delta$'s and therefore of the Yukawa couplings. This estimation agrees with other estimations, for instance, \cite{Georgi:1978xz,Goffin:1979mt,Dine:2015jga}. The parameter $\lambda_5$ can not be too small because of the collider limit on the CP-odd state's mass $A$. Indeed $m_A^2 = 2 \lambda_5 (v_1^2+v_2^2)$ from which it follows 
$\lambda_5\gtrsim 0.02$ from $m_A \gtrsim 50$\,GeV.

 \section{Experimental limits on strong CP parameter}

To test this expression, it is necessary to compare it with experimental measurements. To obtain the relation between the measured neutron electric dipole moment (EDM) \cite{Pendlebury:2015lrz} and $\overline{\theta}$, one must compute the corresponding contribution from the expectation value $\langle n|J_{\mu}| n \rangle$, which requires an estimation of non-perturbative effects. The first computation of this relation was performed in Ref.~\cite{Crewther:1979pi} using chiral perturbation theory. In this approach, one-loop pion penguin diagrams with the emission of photons between neutron states were calculated. In contrast, Ref.~\cite{Pospelov:1999mv} used the operator product expansion and sum rules. More recently, ab initio calculations from lattice QCD have also been used to compute the relation \cite{Guo:2015tla}. However, despite significant effort, the relation is still known with considerable uncertainty, with at least a $\sim 20\%$ error. Furthermore, Ref.~\cite{Abramczyk:2017oxr} pointed out that the form factors used in Ref.~\cite{Guo:2015tla} were incorrect and provided the necessary corrections. Recently, using the gradient flow in the lattice the relation was recalculated
\cite{Liang:2023jfj,Dragos:2019oxn}, finding agreement with the sum rule method. 
Additionally, a recent re-calculation of the sum-rule approach was conducted in Ref.~\cite{Ema:2024vfn}. In that work, the relation between $\overline{\theta}$ and the neutron EDM was obtained using a different linear combination of the two independent nucleon currents contributing to the neutron EDM. The conclusion was that extracting numerically reliable results with this new linear combination is challenging. Consequently, there is significant uncertainty regarding the relation, which weakens the present constraints on new physics models derived from neutron EDM measurements. In this work, we assume that the experimental constraints correspond to $\overline{\theta}$ centered at zero with an error of $5\times10^{-10}$ at $1~\sigma$. It is worth noting the importance of the experimental efforts to measure a signal on the strong CP problem not only of the neutron EDM but also in time reversal invariance violation (TRIV)  neutron-nucleus interaction experiments that will test $\theta$ in the same order of magnitude as the neutron EDM experiments.~\cite{Fadeev:2019bwc,Schaper:2020akw,Kitaguchi:2022sij}.

 \section{Results}

As in the SM, the mass matrices $M_{u,d}^0$ are diagonalized using a bi-unitary transformation ${V_L^{u,d}}^\dagger M_{u,d}^0 V_R^{u,d}$, with the implication that the charged current interactions are not diagonal in the quark mass basis. The mixing is described by the Cabibbo–Kobayashi–Maskawa (CKM) matrix $V_{\rm CKM} \equiv {V_L^{u}}^\dagger   V_L^{d}$. In our model (similar to observations found by Georgi in Ref.~\cite{Georgi:1978xz} for the two-flavor case) the mass matrices satisfy $\operatorname{Arg\,Det}(M_u^0\,M_d^0) = 0$, but $\operatorname{Det} [ M_u^0  {M_u^0}^\dagger, M_d^0  {M_d^0}^\dagger ] \neq 0$, which leaves room to agree with the Jarlskog determinant. 

Generally, two Higgs doublet models yield flavor-changing neutral current (FCNC) at the tree level. This can readily be seen in the Higgs basis \cite{PhysRevD.19.945,Georgi:1978ri,Botella:1994cs}, where one rotates $\phi_1$ and $\phi_2$ to $H_1$ and $H_2$ in such a way that only $H_1$ acquires a vev. The quark masses are therefore given by $H_1$, and the interaction between this Higgs and the quarks is diagonal in the quark mass basis. In contrast, the Yukawa structure that couples $H_2$ to the quarks is not necessarily diagonal in the quark mass basis. This would lead to flavor-changing neutral interactions mediated by $H_2$. Typically, a final rotation is needed to rotate the scalar fields containing $H_1$ and $H_2$ to the scalar mass basis, yielding flavor-changing couplings between all the neutral scalars and the quarks. Here, we assume that the lightest scalar is the Higgs field and that it makes the strongest contribution to a flavor-changing process.

In the quark flavor basis, the Yukawa interaction with $H_2$ is
\begin{eqnarray}
N_u^0&=&\frac{1}{\sqrt{2}}
\left( 
\begin{array}{ccc}
-s_\beta Y^u_{11} & -s_\beta Y^u_{12}& -s_\beta Y^u_{13}\\
-s_\beta Y^u_{21} & -s_\beta Y^u_{22}& -s_\beta Y^u_{23}\\
Y^u_{31} c_\beta e^{i\alpha} & Y^u_{32}  c_\beta e^{i\alpha} & Y^u_{33} c_\beta  e^{i\alpha} 
\end{array}
\right)\,,\nonumber\\
&& \label{Nud}\\
N_d^0&=&\frac{1}{\sqrt{2}}
\left( 
\begin{array}{ccc}
-s_\beta Y^d_{11} &-s_\beta Y^d_{12}& -s_\beta Y^d_{13}\\
-s_\beta Y^d_{21} &-s_\beta Y^d_{22}& -s_\beta Y^d_{23}\\
Y^d_{31} c_\beta e^{-i\alpha} & Y^d_{32}  c_\beta e^{-i\alpha} & Y^d_{33} c_\beta  e^{-i\alpha} 
\end{array}
\right)\nonumber
\end{eqnarray}
By applying the same bi-unitary transformation that leads to the diagonalization of the mass matrices, one can then obtain the FCNC. In Ref. \cite{Alves:2017xmk}, it was shown that matrices with the structure of Eqs. \ref{Nud} lead to controlled flavour-changing contributions. Their analysis was based on the results of Ref. \cite{Blankenburg:2012ex}, which computed the bounds on the quark's effective flavor-changing couplings to a scalar of mass $125$ GeV from low-energy flavor-changing constraints, like meson mixing. We use their notation and write the interaction in the mass basis as
\begin{equation}
\mathcal{L}_{eff}=\sum_{i,j=d,s,b\,i\neq j}C^d_{ij}\overline{d}^{i}_{L}d^{j}_{R}h+\sum_{i,j=u,c,t\,i\neq j}C^u_{ij}\overline{u}^{i}_{L}u^{j}_{R}h
\end{equation}
where $C^{d,u}_{ij}=V^{d,u \dagger}_L N_{d,u}^0 V^{d,u}_L$ for $i,j=d,s,b$ and $i,j=u,c,t$ respectively.

To determine whether the model is viable, we constructed a $\chi^2$, which includes the quark masses, the CKM elements \cite{ParticleDataGroup:2024pth} shown in Tab. \ref{tab:observables_CKM}, and the FCNC constraints from Ref. \cite{Blankenburg:2012ex} shown in Tab. \ref{tab:FCNC constraints}. Therefore, the total number of constraints is 35: six quark masses, three mixing angles, the Jarlskog determinant, 24 low energy flavor changing constraints from \cite{Blankenburg:2012ex}, and $\overline{\theta}$. While the number of free parameters is 20 (18 real yukawas, the CP breaking phase, and $t_\beta$). We note that we fix the value of the parameter $\lambda_5$ that enters in eq.(\ref{loop}) to be $\lambda_5=1$.  The $\chi^2$ is a complicated function of the Yukawas with many local minima. Nevertheless, finding the global minima is unnecessary since we must show that a neighborhood exists where the model is compatible with the experiments.

Just as in the case of the Standard Model, constraints from perturbative unitarity can impose strong bounds on the scalar masses of the theory. In our calculations, we assumed that the mass of the heaviest scalar is around $1\,$TeV. We take this value since it is the largest that can be achieved without fine-tuning the $\mathbb{Z}_2$ soft-breaking term, as shown in Ref. \cite{Nebot:2019qvr}. 
In addition, we comply with the constraints from the Peskin–Takeuchi (S, T, U) parameters, perturbativity, and the boundedness from below of the potential, using a value of $t_\beta < 5$, as shown in Ref.~\cite{Nebot:2019qvr}.

It is indeed possible to find values of the Yukawas that agree with the experimental constraints. The benchmark point is given in Tab. \ref{tab:benchmark}, while the tensions with the experimental constraints can be glanced from Tabs. \ref{tab:observables_CKM} and \ref{tab:FCNC constraints}.
 One must be aware that there is fine-tuning since relatively small variations $\sim 1\%$ yield a drastic increase of the tension. For example, changing the first $Y^{u}_{11}$ by this amount changes the $\chi^2$ by four orders of magnitude. This is a typical situation in two Higgs doublet models.

Other constraints, like $\Gamma(t\rightarrow hq)$ measured at ATLAS \cite{ATLAS:2014lfm,ATLAS:2023ujo,ATLAS:2018jqi} and CMS \cite{CMS:2014jkv,CMS:2016obj} also impose constraints on the flavor violating couplings, although looser than the ones from indirect low energy flavor changing process \cite{Nebot:2019qvr}. 

 \section{Conclusions}
In this letter, we presented a model that addresses the strong CP problem through spontaneous symmetry breaking, drawing inspiration from the model proposed by Georgi \cite{Georgi:1978xz}. By treating CP as a symmetry of the Lagrangian, the model naturally sets the topological term $\theta$ to zero. After symmetry breaking, corrections to this term make it non-zero, but its value remains small. This minimal model requires one additional Higgs doublet with a  $\mathbb{Z}_2$ symmetry which is softly broken. The $\mathbb{Z}_2$ symmetry plays a crucial role in ensuring that the texture of the fermion matrices leads to controlled flavor-changing neutral currents (FCNCs). Despite fine-tuning required in the Yukawa sector, the model remains viable for solving the CP problem, successfully satisfying current experimental constraints while maintaining theoretical consistency. A more detailed analysis of constraints from colliders and a complete global fit, including the scalar sector, is left for future work.

\begin{table}[h]
    \centering
    \begin{tabular}{|c|c|c|c|}
        \hline
        \textbf{Observable} & \textbf{Optimal point} & \textbf{Exp Error} & \textbf{Pull} \\ \hline
        $\hat{m}_u(M_W)$ & 1.3 MeV& 0.4 MeV & 0.08 \\ \hline
        $\hat{m}_d(M_W)$ & 2.7 MeV & 0.4 MeV & 0.04 \\ \hline
        $\hat{m}_s(M_W)$ & 54 MeV & 10 MeV & 0.57 \\ \hline
        $\hat{m}_c(M_W)$ & 0.631 GeV & 0.009 GeV & 0.05 \\ \hline
        $\hat{m}_b(M_W)$ & 2.883 GeV & 0.008 GeV & 0.04 \\ \hline
        $\hat{m}_t(M_W)$ & 173.1 GeV & 0.6 GeV & 0.04 \\ \hline
        $|V_{ud}|$ & 0.97435 & 0.00016 & 0.01 \\ \hline
        $|V_{us}|$ & 0.22501 & 0.00068 & 0.00 \\ \hline
        $|V_{ub}|$ & 0.003732 & 0.000090 & 0.19 \\ \hline
        $|V_{cd}|$ & 0.22487 & 0.00068 & -0.02 \\ \hline
        $|V_{cs}|$ & 0.97349 &  0.00016 & 0.02 \\ \hline
        $|V_{cb}|$ & 0.04183 & 0.0008 & -0.02 \\ \hline
        $|V_{td}|$ & 0.00858 & 0.00019 & 0.05 \\ \hline
        $|V_{ts}|$ & 0.041110 & 0.00077 & -0.03 \\ \hline
        $|V_{tb}|$ & 0.999118 & 0.000034 & 0.02 \\ \hline
        $J$ & 0.000031 & 0.000001 & 0.17 \\ \hline
        $\overline{\theta}$ & $-8.3\times 10^{-10}$ & $5\times10^{-10}$ & -1.7 \\ \hline
    \end{tabular}
    \caption{Quark masses and CKM elements. For asymmetric errors given in \cite{ParticleDataGroup:2024pth} we take as reference the largest one to compute the pulls (the pull is defined as $\left(\mathcal{O}_{exp}-\mathcal{O}_{model}\right)/ \delta O_{exp}$).  Given the good agreement, taking one or the other gives irrelevant changes in the overall agreement. We do show the experimental value used as can trivially be deduced from the pull.}
    \label{tab:observables_CKM}
\end{table}

\begin{table}[h]
    \centering
    \begin{tabular}{|c|c|c|}
        \hline
        \textbf{Observable} & \textbf{Bound 95\%} & \textbf{Model/Bound} \\ \hline
        $|C_{sd}C^{*}_{ds}|$ & \(1.1 \times 10^{-10}\) & \(8 \times 10^{-8}\) \\ \hline
        $|C^2_{ds}|$ & \(2.2 \times 10^{-10}\) & \(8 \times 10^{-7}\) \\ \hline
        $|C^2_{sd}|$ & \(2.2 \times 10^{-10}\) & \(2 \times 10^{-9}\) \\ \hline
        \text{Im}($C_{sd}C^{*}_{ds}$) & \(4.1 \times 10^{-13}\) & \(3 \times 10^{-14}\) \\ \hline
        \text{Im}($C^2_{ds}$) & \(8 \times 10^{-13}\) & \(-1 \times 10^{-14}\) \\ \hline
        \text{Im}($C^2_{sd}$) & \(8 \times 10^{-13}\) & \(-2 \times 10^{-15}\) \\ \hline
        $|C_{cu}C^{*}_{uc}|$ & \(9 \times 10^{-10}\) & \(2 \times 10^{-7}\) \\ \hline
        $|C^2_{uc}|$ & \(1.4 \times 10^{-9}\) & \(8 \times 10^{-5}\) \\ \hline
        $|C^2_{cu}|$ & \(1.4 \times 10^{-9}\) & \(3 \times 10^{-10}\) \\ \hline
        \text{Im}($C_{cu}C^{*}_{uc}$) & \(1.7 \times 10^{-10}\) & \(2 \times 10^{-11}\) \\ \hline
        \text{Im}($C^2_{uc}$) & \(2.5 \times 10^{-10}\) & \(4 \times 10^{-12}\) \\ \hline
        \text{Im}($C^2_{cu}$) & \(2.5 \times 10^{-10}\) & \(-6 \times 10^{-14}\) \\ \hline
        $|C_{bd}C^{*}_{db}|$ & \(9 \times 10^{-9}\) & \(9 \times 10^{-4}\) \\ \hline
        $|C^2_{db}|$ & \(1 \times 10^{-8}\) & \(0.8\) \\ \hline
        $|C^2_{bd}|$ & \(1 \times 10^{-8}\) & \(7 \times 10^{-7}\) \\ \hline
        \text{Im}($C_{bd}C^{*}_{db}$) & \(2.7 \times 10^{-9}\) & \(6 \times 10^{-16}\) \\ \hline
        \text{Im}($C^2_{db}$) & \(3 \times 10^{-9}\) & \(2 \times 10^{-14}\) \\ \hline
        \text{Im}($C^2_{bd}$) & \(3 \times 10^{-9}\) & \(1 \times 10^{-18}\) \\ \hline
        $|C_{bs}C^{*}_{sb}|$ & \(2 \times 10^{-7}\) & \(0.02\) \\ \hline
        $|C^2_{sb}|$ & \(2.2 \times 10^{-7}\) & \(0.8\) \\ \hline
        $|C^2_{bs}|$ & \(2.2 \times 10^{-7}\) & \(2 \times 10^{-4}\) \\ \hline
        \text{Im}($C_{bs}C^{*}_{sb}$) & \(2 \times 10^{-7}\) & \(-2 \times 10^{-17}\) \\ \hline
        \text{Im}($C^2_{sb}$) & \(2.2 \times 10^{-7}\) & \(1 \times 10^{-15}\) \\ \hline
        \text{Im}($C^2_{bs}$) & \(2.2 \times 10^{-7}\) & \(1 \times 10^{-18}\) \\
        \hline
    \end{tabular}
    \caption{Comparison of FCNC 95\% confidence range to the benchmark point prediction. If the ratio is bigger than one there is tension of 2$\sigma$.}
    \label{tab:FCNC constraints}
\end{table}

\begin{table}[h]
    \centering
    \begin{tabular}{|c|c|}
        \hline
        \textbf{Parameter} & \textbf{Value} \\
        \hline
        $\tan\beta$ & 4.14177 \\
        \hline
        $Y^{u}_{11}$ & $+6.99737\times10^{-2}$ \\
        \hline
        $Y^{u}_{12}$ & $-7.33074\times10^{-2}$ \\
        \hline
        $Y^{u}_{13}$ & $-7.11714\times10^{-2}$ \\
        \hline
        $Y^{u}_{21}$ & $-6.87086\times10^{-2}$ \\
        \hline
        $Y^{u}_{22}$ & $+7.22767\times10^{-2}$ \\
        \hline
        $Y^{u}_{23}$ & $+6.96334\times10^{-2}$ \\
        \hline
        $Y^{u}_{31}$ & $+5.63088\times10^{-1}$ \\
        \hline
        $Y^{u}_{32}$ & $-6.72942\times10^{-1}$ \\
        \hline
        $Y^{u}_{33}$ & $-5.25847\times10^{-1}$ \\
        \hline
        $Y^{d}_{11}$ & $+8.51142 \times 10^{-6}$ \\
        \hline
        $Y^{d}_{12}$ & $-6.09872\times10^{-4}$ \\
        \hline
        $Y^{d}_{13}$ & $+3.37296\times10^{-4}$ \\
        \hline
        $Y^{d}_{21}$ & $+1.81040\times10^{-4}$ \\
        \hline
        $Y^{d}_{22}$ & $+1.05304\times10^{-3}$ \\
        \hline
        $Y^{d}_{23}$ & $-4.03567\times10^{-4}$ \\
        \hline
        $Y^{d}_{31}$ & $+1.58555\times10^{-2}$ \\
        \hline
        $Y^{d}_{32}$ & $+6.25660\times10^{-3}$ \\
        \hline
        $Y^{d}_{33}$ & $-2.00748\times10^{-4}$ \\
        \hline
        $\alpha$ & -2.37994 \\
        \hline
    \end{tabular}
    \caption{Benchmark Point used to obtain the results of Tabs. \ref{tab:observables_CKM} and \ref{tab:FCNC constraints}. }
    \label{tab:benchmark}
\end{table}

\section{Acknowledgements}
The authors would like to thank the Libertad Barr\'on-Palos and Jens Erler for useful comments and discussions. This work is supported by the DGAPA UNAM grant PAPIIT IN111625. EP is grateful for the support of PASPA-DGAPA, UNAM for a sabbatical leave, and Fundación Marcos Moshinsky. RFH gratefully acknowledge the Physics Institute at UNAM in Mexico City for their generous hospitality, which provided an excellent environment for conducting this work.

%\appendix
\bibliography{apssamp}

%%\cite{Svrcek:2006yi}
%\bibitem{Svrcek:2006yi}
%P.~Svrcek and E.~Witten,
%%``Axions In String Theory,''
%JHEP \textbf{06} (2006), 051
%doi:10.1088/1126-6708/2006/06/051
%[arXiv:hep-th/0605206 [hep-th]].

\end{document}